\documentclass[journal]{IEEEtran}

\usepackage{blindtext, graphicx}

\ifCLASSINFOpdf
\else
\fi
\begin{document}
%
\title{4-Bit High-Speed Binary Ling Adder}



%
\author{Projjal~Gupta,~\IEEEmembership{Member,~IEEE}

Electronics and Communication Engineering

SRM Institute of Science and Technology, Kattankulathur

Email : projjal.gupta@nextech.io

}

\maketitle


\begin{abstract}
Binary addition is one of the most primitive and most
commonly used applications in computer arithmetic. A
large variety of algorithms and implementations have been
proposed for binary addition. Huey Ling proposed a simpler form of
CLA equations which rely on adjacent pair bits (ai, bi) and
(ai−1, bi−1). Along with bit generate and bit propagate, we
introduce another prefix bit, the half sum bit. Ling adder increases the speed of n-bit binary addition, which is an upgrade from the existing Carry-Look-Ahead adder. Several variants of the carry look-ahead equations, like Ling carries, have been presented that simplify carry computation and can lead to faster structures. Ling adders, make use of Ling carry and propagate bits, in order to calculate the sum bit. As a result, dependency on the previous bit addition is reduced; that is, ripple effect is lowered. This paper provides a comparative study on the implementation of the above mentioned high-speed adders.
\end{abstract}

\begin{IEEEkeywords}
Ling Adder, High Speed Binary Adder, Binary Addition.
\end{IEEEkeywords}

%
\IEEEpeerreviewmaketitle

\section{Introduction}

The family of Ling adders is a particularly fast adder and is designed using H. Ling's equations and generally implemented in BiCMOS. It is an upgrade to the already existing Carry-Look-Ahead Adders and is mathematically faster, as it requires lesser steps for the computation of a sum. The circuit of a Ling adder is particularly more complex, and is less favourable for use in VLSI systems due to its complexity and it requires far more extra components than traditional systems. The circuit is divided into 4 parts, which can be denoted by H. Ling's equations.

\section{Analysis of Ling's Equation}
\subsection{Initial Generation of Bits}
Ling Adders require to form the bit generate and bit propogate that are used in the regular Carry look ahead adders. It is denoted by the 3 symbols $g_{i}$, $p_{i}$ and $d_{i}$.\\
The generate and propagate bits follow CLA so they can be denoted as,
\[g_{i} = a_{i} \cdot b_{i}\] 
\[p_{i} = a_{i} + b_{i}\]

However, Ling adder requires an extra half bit term which later on simplifies the circuit design, while increasing the overall efficiency of the adder. This half bit generate is denoted by $d_{i}$ and can be mathematically shown by,
\[d_{i} = a_{i} \oplus b_{i}\]

The above mentioned Generate Bit $g_{i}$ and Propagate Bit $p_{i}$ are used further to derive the Ling Generates, which are terms that will go on to simplify the final equation. This is particularly important because these generates will form the base of the Ling adder circuit design. These are denoted by $G_{i}^*$ and $P_{i}^*$ \\

{\includegraphics[width=3.25in,height=5in,keepaspectratio]{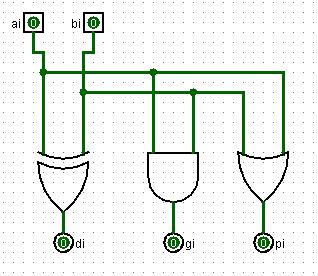}}

\subsection{The CLA Basis of Ling's Equations}
The CLA depends upon the carry out term of the previous for the new carry terms.
\[c_{i+i} = g_{i} + p_{i}\cdot c_{i}\]
which is similar to the Simple Ripple Adder which uses the carry output of the preceding data bits for forward addition. \\

Similarly based on the above concept, Ling created a new theoretical carry generate which he denoted by H. This is used later on in the Adder to generate the sum $S_{i}$. The term H is given by,
\[H_{i} = c_{i} + c_{i-1}\] where,
\[c_{i} = H_{i}\cdot p_{i}\]

Introduction of ling carry $H_{i}$ is one of the major reasons why Ling Adder is a fast yet complex adder. Use of Ling carry equations decreases the number of boolean terms during its operations, but increases the design complexity.  

\subsection{Ling Generate and Propagate}
Ling proposed the use of Ling Propagate and Ling Generate to simplify the operations of the Ling adder. It is very important, as this is the first step where we can see how the terms are generated by using the $i^{th}$ and $(i-1)^{th}$ terms. These terms can be derived by
\[G_{i}^{*} = g_{i} + g_{i-1}\] and
\[P_{i}^{*} = p_{i} \cdot p_{i-1}\]

Ling generate and propagate terms are used to calculate the Ling carry term $H$. Later on in the Adder design, the sum terms are directly influenced by the all the Ling terms.

{\includegraphics[width=3.25in,height=5in,keepaspectratio]{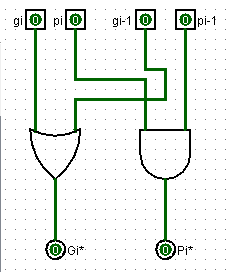}}

\subsection{Ling Sum Term}
The final sum term for the $i^{th}$ pair terms of a and b are devised by following Ling sum equations, which take in lesser number of inputs, and hence decrease the lag in the system.\\

If we assume that all input gates have only two inputs, we can see that calculation of CLA carry $C$ requires 5 logic levels, whereas that for ling carry $H$ requires only four. Although the computation of carry is simplified, calculation of the sum bits using Ling carries is much more complicated. The sum bit, when calculated by using traditional carry, is given to be,
\[s_{i} = d_{i} \oplus c_{i-1}\]

We note that we require to use both the carry output and half-bit term from the first operation block.
\[c_{i} = H_{i}\cdot p_{i}\] Using the above term in the Ling Sum Equation,
\[s_{i} = d_{i} \oplus p_{i-1}\cdot H_{i-1}\] on break down,
\[s_{i} = H'_{i-1}\oplus d_{i} + H_{i-1}(d_{i} \oplus p_{i-1})\]

{\includegraphics[width=3.25in,height=5in,keepaspectratio]{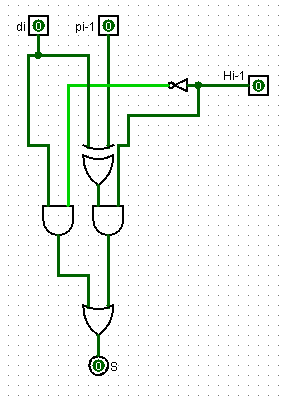}}

Hence, the output value for $a_{i} + b_{i}$ is given by $s_{i}$ and $c_{i}$.

\section{Ling Carry Equation}
\subsection{General expansion and Substitution}
Earlier in the CLA basis subsection of Ling Equation analysis, we came across 2 equations,
\[H_{i} = c_{i} + c_{i-1}\] and,
\[c_{i} = H_{i}\cdot p_{i}\]
Since we know that,
\[c_{i+1} = g_{i} + p_{i}\cdot c_{i}\] Thus we can write the Carry Output as,
\[H_{i} = g_{i}+g_{i-1}+p_{i-1}\cdot g_{i-1}+p_{i-1}\cdot p_{i-2} \cdot g_{i-2}+.....\]\[+ p_{i-1}\cdot p_{i-2}\cdot p_{i-1}\cdot.....\cdot p_{1}\cdot p_{0}\cdot g_{0}\]

But we know that,
\[G_{i}^{*} = g_{i} + g_{i-1}\]
\[P_{i}^{*} = p_{i} \cdot p_{i-1}\]

Thus we can simplify the $H$ equations to Ling generate-propagate terms.

\subsection{Application in 4-Bit System}
In a 4-bit adder design, we require the terms $H_{3}$, $H_{2}$, $H_{1}$ and $H_{0}$.

From the Ling generate-propagate equations and the expanded Ling Carry equation in the previous subsection, we can write the 4 terms as,
\[H_{3} = G_{3}+P_{2}\cdot G_{1}\]
\[H_{2} = G_{2}+P_{1}\cdot G_{0}\]
\[H_{1} = G_{1}\]
\[H_{0} = G_{0}\]

It is noted that the complexity of the system will increase with increase of Input terms.

\section{Logic design of 4-Bit Ling Adder}

From all the above sections and designs, we can design the 4 bit Ling Adder.\\

{\includegraphics[width=6.9in,height=5.25in]{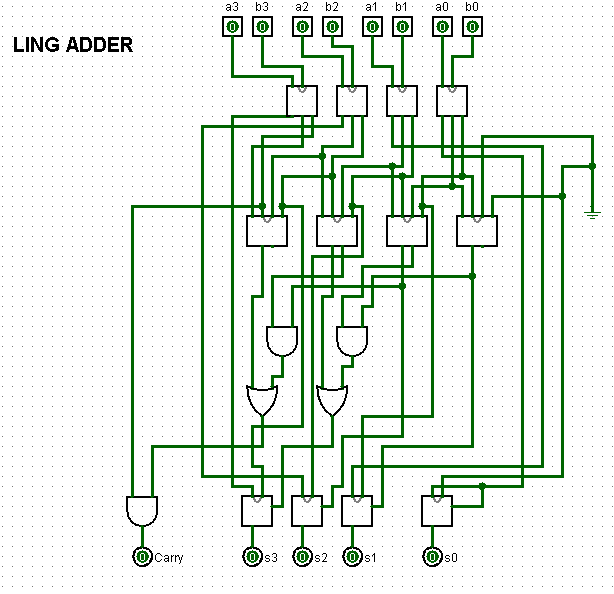}}

As per the design, The flowing outputs are passed through basic OR and AND gates to satisfy $H$ equations.The carry is safely calculated as $c_{4}$
\[c_{4} = H_{4}\cdot p_{4}\]
Similarly, each block present in the Logic Diagram represents an operation step described in each subsection of the logic analysis. \\
The use of free gates in the circuit represent the operations used to calculate $H_{i}$. The initial $g_{-1}$ and $p_{-1}$ dont exist during the case $i = 0$ and hence they are taken as logical $false$ or $zero$ value by grounding them. \\
Effectively, the overall circuit follows the final equation 
\[s = a + b\] and generates a carry term in-case the overall sum exceeds the 4-bit output range.

\vspace*{40\baselineskip}

\bigbreak

\section{PCB and CAD Design}
The system can be designed in real time by the use of actual logic gate ICs belonging to the 74xx family. These ICs usually consist of 16 (DIL16) or 14 (DIL14) pins, and require low power. From the above Logisim design of the Ling Adder, we can start designing the same circuit on any EDA or CAD software. Due to its high complexity, the circuit has to be designed on the both sides of a pcb and uses multiple vias for the on-board connections.

{\includegraphics[width=4.5in,height=6in,keepaspectratio]{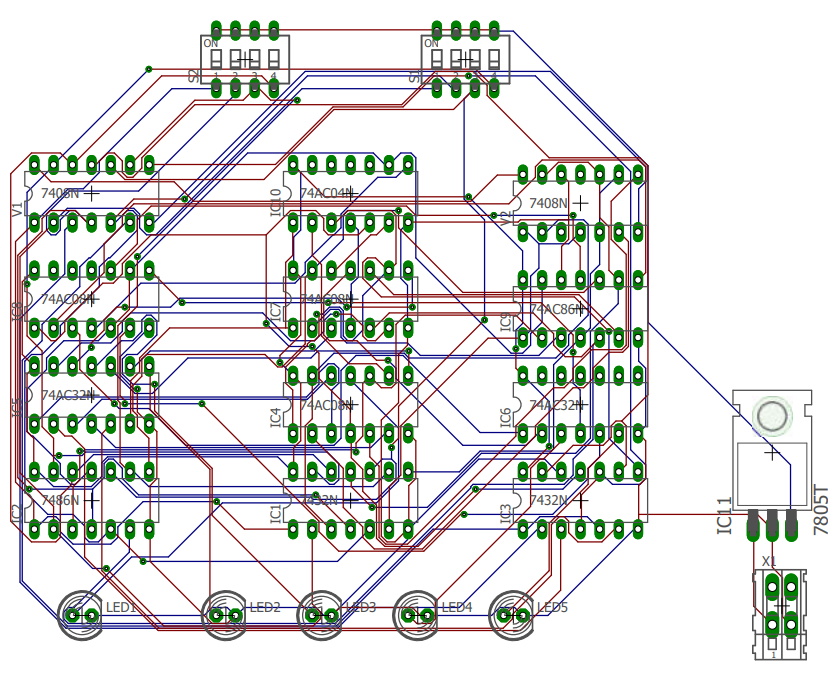}}

\section{Conclusion}
Hence, a basic 4-Bit Ling adder circuit was designed according to Huey Ling's equations. In 4-Bit arithmetic system, the CLA requires 5 terms, whereas the Ling adder requires a maximum input of 4 terms, thereby decreasing the time required for computation. When this adder is cascaded for higher number of Bit input terms, the CLA will come across an increase in operation times. But in a ling adder, this time increase would be much lesser than the other binary adders.\\


%

\section*{Acknowledgment}

The authors would like to thank Mr. AVM Manikandan (Asst. Professor, ECE Dept.) for his teachings, his support and guidance, under which the project was successfully completed.

\ifCLASSOPTIONcaptionsoff
  \newpage
\fi

\end{document}